\documentclass[a4paper,fleqn]{cas-sc}
\usepackage[numbers,sort&compress]{natbib}
\usepackage{algorithm}
\usepackage{algpseudocode}
\usepackage{adjustbox}

\usepackage{multirow}
\usepackage{makecell}
\usepackage[shortlabels]{enumitem}

\def\tsc#1{\csdef{#1}{\textsc{\lowercase{#1}}\xspace}}
\tsc{WGM}
\tsc{QE}
\tsc{EP}
\tsc{PMS}
\tsc{BEC}
\tsc{DE}

\begin{document}
\let\WriteBookmarks\relax
\let\printorcid\relax
\def\floatpagepagefraction{1}
\def\textpagefraction{.001}
\shorttitle{Day-Ahead Electricity Price Forecasting}

\shortauthors{Wang et~al.}
\title [mode = title]{Day-Ahead Electricity Price Forecasting Using a Multivariate Group Lasso Method}

\author[1]{Keyi Wang}

\author[1]{Jiaxiang Ji}

\author[2]{Mahan Mansouri}

\author[1,3]{Ahmed Aziz Ezzat}
\cormark[1]
\ead{aziz.ezzat@rutgers.edu}

\affiliation[1]{
organization={Rutgers, The State University of New Jersey},
city={Piscataway},
postcode={08854},
state={New Jersey},
country={United States of America}
}

\affiliation[2]{
organization={PG\&E, Pacific Gas \& Electric Company},
city={Oakland},
postcode={94612},
state={California},
country={United States of America}
}

\affiliation[3]{
organization={University College, Korea University},
city={Seoul},
country={South Korea}
}

\cortext[cor1]{Corresponding author}

\begin{abstract}
Electricity price signals in modern power systems exhibit complex dependence structures that render forecasting inherently challenging. Our analysis of real-world electricity pricing signals 
reveals complex temporal group effects, whereby the influence of explanatory variables on electricity prices persists across consecutive blocks of time due to underlying economic, system, and operational drivers. In response, we propose a multivariate statistical method based on a Group Lasso formulation to jointly forecast the vector of day-ahead electricity prices ($h = 1, ..., 24$), by leveraging multi-feature temporal group effects. Our approach is evaluated on two full years of electricity prices from the California Independent System Operator (CAISO), demonstrating considerable improvements in point and probabilistic forecast metrics compared to a wide array of statistical and deep learning methods. Empirical analyses confirm the effectiveness of the proposed approach in modeling realistic group effects, maintaining both interpretability and low computational complexity. 
When retrospectively evaluated on test data from a recent international electricity price forecasting challenge, the proposed method ranked in second place, despite having access to significantly less information than competing approaches. Finally, the proposed method is independently validated against two operational electricity price forecasting systems in CAISO, demonstrating competitive predictive performance and practical relevance.
\end{abstract}


\maketitle

\section{Introduction}
In modern power systems, accurate short-term electricity price forecasts are critical inputs to reliable and economically efficient decision-making by system operators and electricity market participants. For system operators, accurate day-ahead forecasts of pricing signals support secure system balancing and market clearing, while for market participants, they inform bidding and procurement strategies, production scheduling, and asset management across electricity market operations \citep{fan2014min, kwon2016optimal, mazzi2017price, wang2026marketsignals}. Given the substantial economic consequences associated with forecast errors, effective short-term electricity price forecasting (EPF) has become an essential component of modern power system operations and market participation.


There is an extensive literature on short-term EPF models, spanning both statistical and machine learning (ML) paradigms \citep{Weron2014EPF, IISE_PGE}. The vast majority of statistical methods for EPF are rooted in time-series and regression models which explicitly encode autocorrelations, seasonal patterns, and dynamic covariate effects. 
Representative examples include dynamic regression and autoregressive-based models \citep{conejo2005forecasting,zareipour2006application}, exponential smoothing approaches \citep{cruz2011effect}, dynamic quantile regression \citep{bunn2016analysis}, and regime-switching models \citep{karakatsani2008intra}. Multivariate and regularized methods have also been proposed, especially in the presence of a large set of exogenous features. These include 
those based on Lasso, Elastic Net, and related variants \citep{uniejewski2016automated,ziel2018day, 9788043,matsumoto2024electricity, mascarenhas2026leveraging}. 

More recently, ML-based approaches have gained substantial attention due to their ability to learn complex drivers of pricing signals from large feature sets. This line of work includes tree-based and ensemble learning models \citep{mei2014random, wang2022online}, kernel-based approaches \citep{chaabane2014novel}, as well as deep neural networks (DNNs), including those based on 
recurrent \citep{xie2018day}, convolutional \citep{lago2018forecasting}, autoencoder-based and generative models \citep{wang2016short,zhang2021predicting}, as well as transformer-based architectures \citep{zhang2023real}.


There are pros and cons for the statistical and ML approaches for short-term EPF. Generally, statistical methods remain prevalent in practice because they are relatively simpler to train, easier to interpret, and have demonstrated competitive predictive performance especially for shorter forecast horizons. 
Among these statistical approaches, we would like to especially shed light on the \textit{Lasso-estimated autoregressive (LEAR) model}, which is a well-established and well-proven statistical method for short-term EPF \citep{uniejewski2016automated}. 
LEAR is a parameter-rich linear regression model for day-ahead EPF which leverages Lasso regularization to account for diverse covariate sets efficiently, including historical price patterns, operational, economic, environmental, and calendar effects. 
Recent large-scale experiments suggest that LEAR's performance is on par with that of deep learning approaches for EPF, with the added advantages of providing interpretable covariate effect explanations (in the form of estimated coefficients), and having 
compute costs that are 
substantially smaller for both model training and evaluation, favoring their adoption for operational forecasting applications in electricity systems \citep{lago2021forecasting}.

Despite these advantages, a key limitation of the LEAR model lies in its treatment of temporal dependence. 
In its standard formulation, LEAR models each hourly price independently, that is, a separate model is trained for each hourly price in the day-ahead horizon. 
Treating each hour as a separate learning task can overlook systematic similarities in how explanatory variables, such as load and generation, influence prices over time. 
 There is an active line of research to effectively model cross-hour dependencies by adopting multivariate formulations, where the hourly day-ahead prices are modeled jointly rather than as independent outputs. A common approach in this context is vector autoregression (VAR) 
\citep{huisman2007hourly,panagiotelis2008bayesian,haldrup2010vector}. Yet, VAR models become challenging to scale with increasing dimensionality, particularly in the presence of large feature sets. 
There are dimension-reduction strategies such as factor-based and reduced-rank structures that have been proposed to capture cross-hour dependence with a smaller set of parameters \citep{garcia2012forecasting, maciejowska2015forecasting, maciejowska2016short}. Another alternative is to design a multi-output DNN architecture with a vectorized output  representing the hourly day-ahead prices in the following day 
\citep{ziel2018day, lago2021forecasting}. 

While the above approaches can model cross-hour dependencies, they do not explicitly enforce a key requirement: consistent predictor selection over time. This is especially relevant in the presence of \textit{temporal group effects,} whereby the explanatory features influence prices over contiguous blocks of time periods rather than at isolated time points, reflecting underlying operational and economic dynamics. These cross-hour group effects arise from the temporal coupling inherent in electricity system operations. For example, load and generation at a given hour affect unit commitment, ramping decisions, and reserve requirements across multiple hours, leading to non-local dependencies in electricity prices. Price formation is therefore naturally governed by inter-temporal operational constraints that induce structured dependencies across consecutive blocks of time periods. 

\begin{figure*}
    \centering
    \includegraphics[width=1\linewidth, trim = 0 0.25cm 0 0]{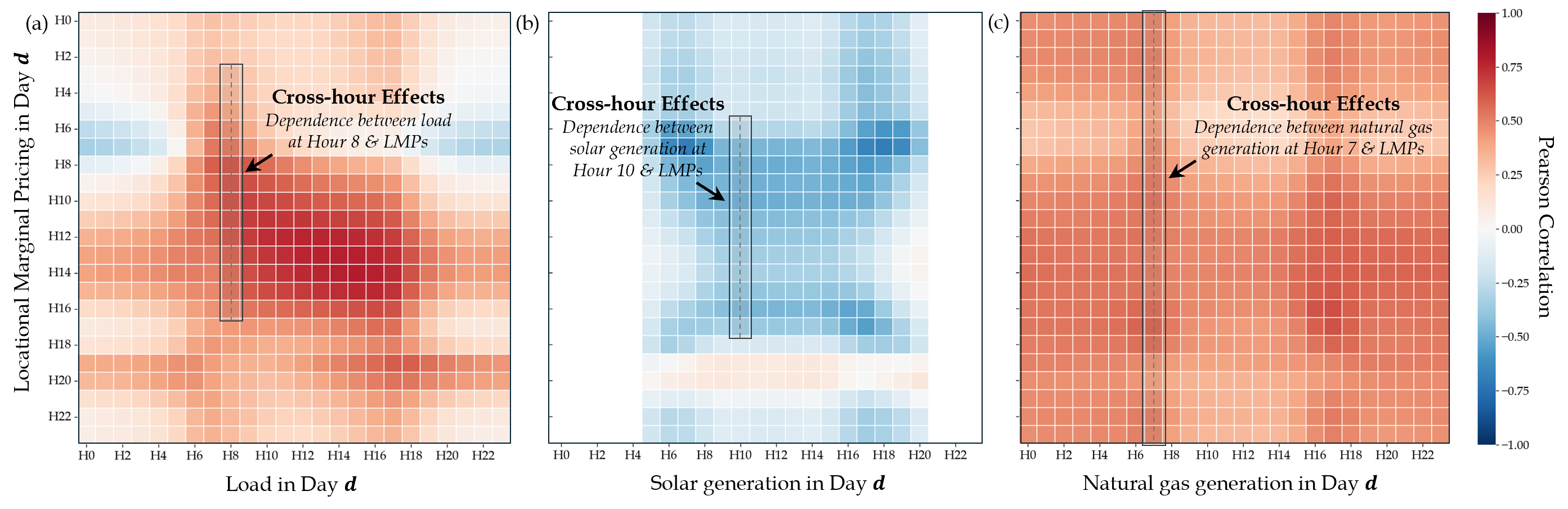}
    \caption{Pearson correlation heatmaps between LMPs and forecasts of regional load (a), solar generation (b), and natural gas generation (c), respectively, during 2025 in CAISO. The plots reveal strong cross-hour (non-local) temporal dependencies between key system variables and LMPs, wherein each column shows the dependence between a single-hour variable (e.g., regional load at hour 8) and the corresponding intraday LMPs. The structured, non-local patterns highlight the importance of modeling temporal group effects in LMPs, which is not captured by hour-wise independent models.}
    \label{fig:correlation}
\end{figure*}

Our analysis of electricity pricing signals from the California Independent System Operator (CAISO) further supports this observation. 
Figure \ref{fig:correlation} shows the strength of Pearson correlations (evaluated using a full year of data from CAISO) between the hourly locational marginal prices (LMP) and the corresponding forecasts of system load (Panel a), solar generation (Panel b), and natural gas generation (Panel c). 
In the absence of cross-hour dependence, one would expect the influence of an hourly explanatory variable (load, solar, or natural gas generation) to be primarily localized to the corresponding hourly LMP, resulting in a correlation structure concentrated along the diagonal of the heatmaps shown in Figure \ref{fig:correlation}.
Instead, the correlation heatmaps exhibit clear vertical band structures, 
suggesting that a given hourly feature is strongly correlated with LMPs across several consecutive hourly blocks. 
For example, morning load levels in Figure \ref{fig:correlation}(a) (hours $6$–$9$) display strong and temporally persistent correlations with LMPs throughout the day. Similarly, daytime solar generation in Figure \ref{fig:correlation}(b) (hours $6$–$19$) shows consistent negative correlations across most time periods, whereas natural gas generation in Figure \ref{fig:correlation}(c) exhibits consistent positive correlations with LMPs across almost all hours, with relatively stronger effects during sunrise and sunset periods (hours $6$-$8$ and $16$–$18$). 


To model these cross-hour group effects, we propose a multivariate statistical model for short-term EPF, which we dub as the ``\underline{C}ross-\underline{IN}fluence-\underline{G}roup-\underline{L}asso-\underline{E}stimated \underline{A}uto-\underline{R}egressive'' model, or in short {CING-LEAR} (pronounced as Shakespeare's King Lear). CING-LEAR generalizes the LEAR model to a multivariate setting and encodes temporal group effects through a group regularization. 
Our primary contributions are summarized as follows:
\begin{itemize}
    \item We propose a multivariate EPF model that accounts for temporal dependencies and cross-hour feature group effects that are prevalent in electricity pricing signals. CING-LEAR extends the LEAR framework to a multi-output setting and is formulated with a Group Lasso regularizer, enabling the joint selection of covariate effects across the full forecast horizon. We posit that this group-wise sparsity structure is more naturally aligned with the inter-temporal dependencies induced by electricity system operations.

    \item We extensively evaluate CING-LEAR on two years of LMPs from CAISO. Large-scale empirical testing suggests considerable improvements in point and probabilistic forecast performance, whereas post hoc feature analysis confirms its effectiveness in capturing realistic group effects.
We also retrospectively evaluate CING-LEAR on a recent international electricity price
forecasting challenge, where it ranks in second place, despite having access to significantly less information than competing approaches. Finally, we independently benchmark it 
against two operational forecasting systems currently deployed by a major U.S. utility where it demonstrates competitive
performance and practical relevance.
\end{itemize}

The remainder of this paper is organized as follows. Section \ref{sec: data} describes the data used in this work. Section \ref{sec : method} details the formulation of CING-LEAR. 
Section \ref{sec:experiment} introduces the experimental setup and evaluation metrics, followed by Section \ref{sec:results} where the results are presented. Finally, Section \ref{sec:conclusion} concludes the paper and highlights future research directions.

\section{Data Description}\label{sec: data}

Four years of day-ahead LMPs at the pricing node NP15SLAK\_5\_N001, along with exogenous information about load, generation, and fuel prices, are extracted from the Grid Status API for the CAISO NP15 region, spanning from March 2021 to December 2025 at hourly resolution \citep{gridstatus}. 
Solar generation is directly available at the NP15 region, whereas other exogenous variables are available at different spatial resolutions. Specifically, load data are obtained from the PG\&E TAC area which closely corresponds to the geographic scope of the NP15 region. Natural gas generation is available only at the system level, while fuel prices are sourced from the BANC region. 

\begin{figure*}
    \centering
    \includegraphics[width=1\linewidth]{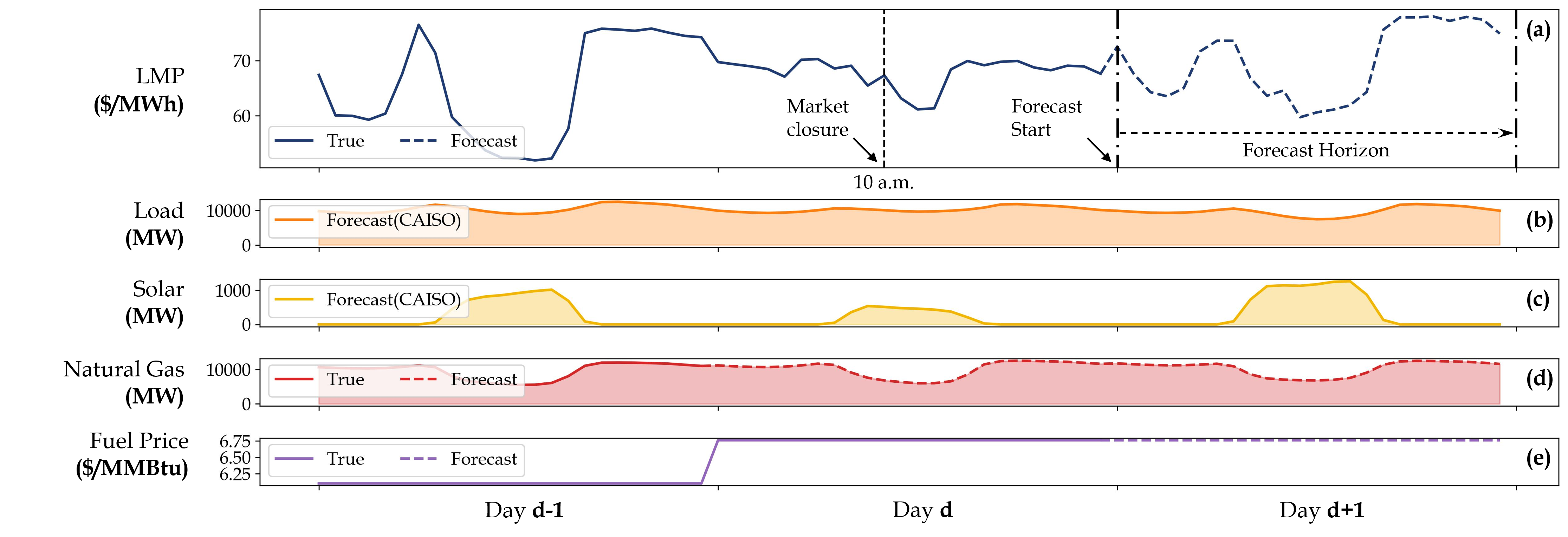}
    \caption{Time series illustration of LMPs and exogenous variables over a three-day horizon. The vertical dashed line in panel (a) denotes the day-ahead market closure time (10:00am), at which the LMP for day $d+1$ is forecasted. Across all panels, solid lines indicate information that are available before the market closure time, while dashed lines represent information that are not available and therefore must be forecasted. Specifically, in panels (a), (d), and (e), solid lines correspond to true values and dashed lines denote forecasted values. Panels (b) and (c) display forecasted values only, as provided by CAISO.}
    \label{fig:DAM}
\end{figure*}

Figure \ref{fig:DAM} illustrates the forecasting setup. Standing at day $d$, we forecast the hourly LMPs of day $d+1$ before the day-ahead market closure time (10:00am). 
To do so, the model is trained using data from the previous three years, including the $d$th day prices since the CAISO day-ahead market results are published at 1:00pm 
on the day prior to the trade day. 
For all exogenous covariates, we use forecasts provided by CAISO when available; otherwise, forecasts are generated using simple in-house models that we develop ourselves. 
%
CAISO publishes its day-ahead forecasts for load and solar generation prior to market closure, released at 9:00am and 8:45am, respectively.  
In contrast, forecasts for natural gas generation and fuel prices are not publicly available, so we generate these forecasts internally, as described next. 

For natural gas generation, we use an ARIMA-X model. 
Let \(g_{d,h}\) denote the natural gas generation at hour \(h\) on day \(d\). Its forecast model is specified as in (\ref{eq:ng}). 
\begin{equation}
\label{eq:ng}
\begin{aligned}
\hat{g}_{d,h}^{\text{ARIMA-X}}
=\phi_0
+ \sum_{i=1}^{p}\phi_i g_{d,h-i}
+ \sum_{j=1}^{q}\theta_j\varepsilon_{d,h-j}
+ \beta_1 g_{d-7,h}
+ \beta_2 \widehat{\ell}_{d,h}
+ \beta_3 \widehat{s}_{d,h} 
+ \beta_4 \widehat{w}_{d,h}
+ \varepsilon_{d,h},
\qquad h=1,\ldots,24.
\end{aligned}
\end{equation}
where \( \{\phi_i\} \) and \( \{\theta_j\} \) denote auto-regressive and moving-average coefficients, respectively. In (\ref{eq:ng}), \(g_{d-7,h}\) denotes the natural gas generation at the same hour one week earlier, whereas
\(\widehat{\ell}_{d,h}\), \(\widehat{s}_{d,h}\) and \(\widehat{w}_{d,h}\) are the 
load, solar and wind forecasts issued by CAISO, respectively, and \( \varepsilon_{d,h} \) is a stochastic error term. 
We experiment with various model configurations and training periods, and conclude that 
an order of $p = q = 1$ with a training length of 28 days yields a sufficiently accurate forecasts. 
Fuel prices, on the other hand, are constant within each day, and are known at the time of forecasting for the $d$th day. We therefore adopt a persistence forecast to predict day-ahead fuel prices. 

Overall, LMPs exhibit a strong positive correlation with fuel prices ($0.74$ on a Pearson correlation scale), and a moderate positive correlation with regional load ($0.36$ on a Pearson correlation scale), while showing a moderate to weak negative correlation with solar generation ($-0.26$ on a Pearson correlation scale). Natural gas generation, which often serves as a complementary resource to variable solar generation as illustrated in Figure \ref{fig:DAM} (c)-(d), is also positively correlated with LMP ($0.44$ on a Pearson correlation scale).

\section{Methodology}
\label{sec : method}
We begin by formally defining important notation in Section \ref{sec:feats}, followed by a brief overview of the LEAR method in Section \ref{sec:lear}. From there, we introduce the building blocks of the proposed approach, CING-LEAR, in Section \ref{sec:CING}.

\subsection{Model inputs and notation} \label{sec:feats}

Let $p_{d,h}$ denote the day-ahead electricity price for day $d$ and hour $h\in\{1,\dots,24\}$. We collect the hourly prices for day $d$ in the vector, $\mathbf{p}_d$ defined as follows: 
\[
\mathbf{p}_d = (p_{d,1},\dots,p_{d,24})^\top \in \mathbb{R}^{24}.
\]

For the exogenous variables, we use the day-ahead forecasts of $K$ variables. For each variable $k\in\{1,\dots,K\}$ we denote the hourly forecast for day $d$ by: 
\[
\mathbf{x}_d^{(k)} = \bigl(x_{d,1}^{(k)},\dots,x_{d,24}^{(k)}\bigr)^\top \in \mathbb{R}^{24}.
\]

The input features used to predict the $24$-dimensional price vector $\mathbf{p}_d$ for day $d$ include the following: 
\begin{itemize}
    \item Historical day-ahead prices of the previous three and seven days, respectively:
    \[
    \mathbf{p}_{d-1},\ \mathbf{p}_{d-2},\ \mathbf{p}_{d-3},\ \mathbf{p}_{d-7}.
    \]
    \item Day-ahead forecasts of the $K$ exogenous variables for days $d$, $d-1$ and $d-7$:
    \[
    \bigl\{\mathbf{x}_d^{(k)},\,\mathbf{x}_{d-1}^{(k)},\,\mathbf{x}_{d-7}^{(k)} : k=1,\dots,K\bigr\}.
    \]
    \item A vector of dummy variables 
    denoting the day-of-week variation: 
    \[
    \mathbf{z}_d = (z_{d,1},\dots,z_{d,7})^\top\in\{0,1\}^7,
    \]
    where $z_{d,j}=1$ if day $d$ is the $j$-th day of the week and $z_{d,j}=0$ otherwise.
\end{itemize}

For notational convenience, we stack all features available on day $d-1$ for predicting $\mathbf{p}_d$ into a single vector:
\[
\begin{aligned}
\mathbf{x}_d
={  \bigl[
    \mathbf{p}_{d-1}^{\top},
    \mathbf{p}_{d-2}^{\top},
    \mathbf{p}_{d-3}^{\top},
    \mathbf{p}_{d-7}^{\top}, 
    \{\mathbf{x}_d^{(k)\top}\}_{k=1}^{K},
    \{\mathbf{x}_{d-1}^{(k)\top}\}_{k=1}^{K}, 
    \{\mathbf{x}_{d-7}^{(k)\top}\}_{k=1}^{K},
    \mathbf{z}_d^{\top}
    \bigr]^{\top}
    \in \mathbb{R}^{M},}
\end{aligned}
\]
where $M$ denotes the total number of input regressors.
Given a dataset of $N$ days $\{d=1,\dots,N\}$, we define the design (or regression) matrix as follows: 
\[
\mathbf{X} = 
\begin{bmatrix}
\mathbf{x}_1^\top\\
\vdots\\
\mathbf{x}_N^\top
\end{bmatrix}\in\mathbb{R}^{N\times M}.
\]

\subsection{Background on the LEAR model for day-ahead electricity price forecasting}
\label{sec:lear}

Here, we briefly review LEAR, a well-established statistical model for short-term EPF 
\citep{uniejewski2016automated}.
LEAR models the electricity price $p_{d,h}$ on day $d$ and hour $h$ using a parameter-rich linear regression formulation as in (\ref{eq:lear}). 
\begin{equation}
\label{eq:lear}
\begin{gathered}
p_{d,h}
= \mathbf{x}_d^{\top}\boldsymbol{\beta}_h
+ \varepsilon_{d,h},
d=1,\ldots,N,\qquad h=1,\ldots,24.
\end{gathered}
\end{equation}
where $\boldsymbol{\beta}_h\in\mathbb{R}^{M}$ denotes an hour-specific vector of regression coefficients, and $\varepsilon_{d,h}$ denotes a zero-mean error term.

LEAR estimates each parameter set $\boldsymbol{\beta}_h$ independently using a Lasso regularization method, which imposes an $\ell_1$ penalty on the coefficient vector in order to perform automatic and sparse feature selection. Specifically, for each hour $h\in\{1,\dots,24\}$, the parameter set $\boldsymbol{\beta}_h$ is estimated using the expression in (\ref{eq:lear_lasso}). 
\begin{equation}
    \widehat{\boldsymbol{\beta}}_h
    =
    \arg\min_{\boldsymbol{\beta}\in\mathbb{R}^{M}}
    \left\{
        \frac{1}{N}\sum_{d=1}^N
        \bigl(p_{d,h}-\mathbf{x}_d^\top\boldsymbol{\beta}\bigr)^2
        + \lambda_h\Vert \boldsymbol{\beta} \Vert_1
    \right\},
    \label{eq:lear_lasso}
\end{equation}

This hour-by-hour Lasso strategy 
produces an independent coefficient vector $\widehat{\boldsymbol{\beta}}_h$ for each hour $h$, enabling the set of selected predictors to vary across the forecast horizon. 

\subsection{The proposed CING-LEAR model for day-ahead electricity price forecasting} \label{sec:CING}

The motivation for CING-LEAR follows directly from the cross-hour feature group effects observed in CAISO's pricing signals\textemdash recall Figure \ref{fig:correlation}. 
The observations made therein motivate us to formulate a multivariate extension of LEAR which specifically incentivizes consistent feature selection across consecutive blocks of time periods, rather than estimating each localized hourly effect independently.

\begin{figure}
    \centering
    \includegraphics[width=0.8\linewidth]{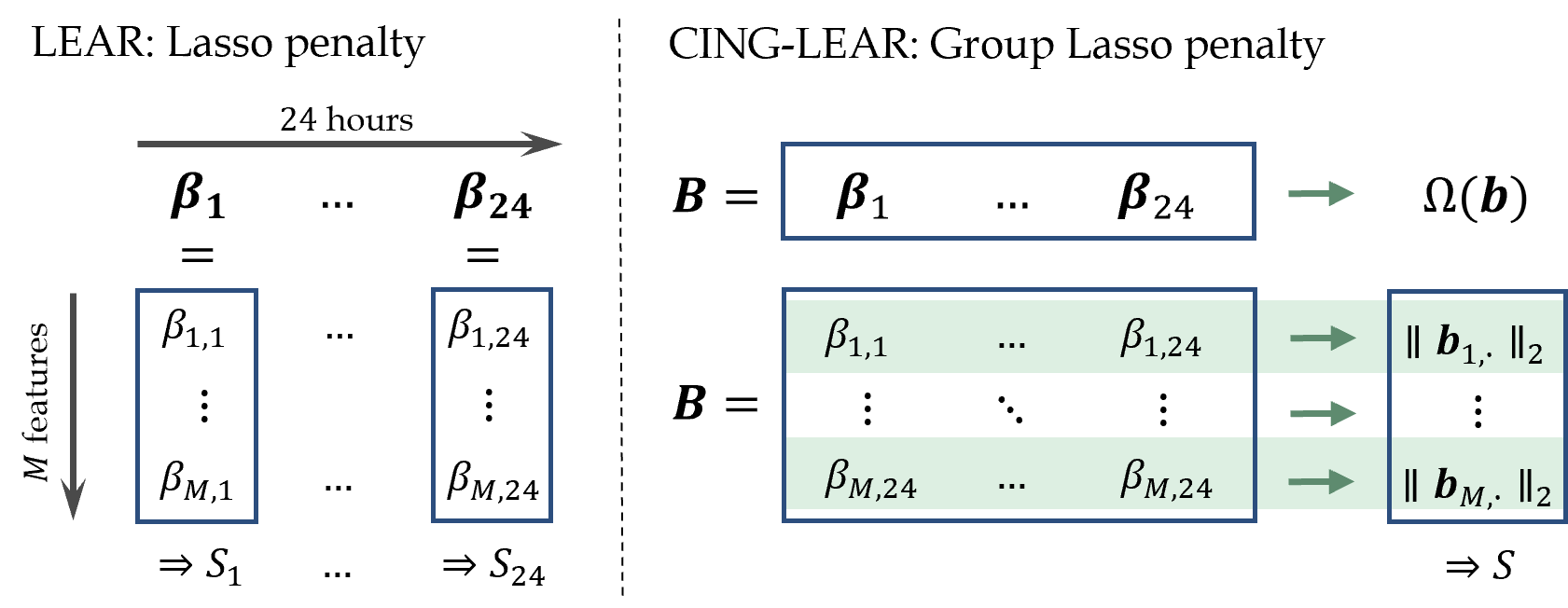}
    \caption{Conceptual difference between the regularization structure in LEAR (left) versus its proposed multivariate extension, CING-LEAR (right).}
    \label{fig:Group_lasso}
\end{figure}

The conceptual difference between LEAR and CING-LEAR can be understood through the coefficient organization scheme shown in Figure \ref{fig:Group_lasso}. LEAR estimates 24 hour-specific coefficient vectors $\{\boldsymbol{\beta}_h\}_{h=1}^{24}$ and applies an $\ell_1$ penalty to each $\boldsymbol{\beta}_h$ independently. This hour-by-hour strategy allows different subsets of predictors to be selected for different hours, which can be flexible but does not explicitly encode the cross-hour dependence implied by Figure \ref{fig:correlation}. In contrast, CING-LEAR organizes the coefficients so that each predictor forms a single group across the 24 hourly outputs. This design directly mirrors the vertical-band structure shown in Figure \ref{fig:correlation}: if a covariate tends to consistently influence the electricity prices across hourly blocks, it is natural to treat its day-ahead hourly coefficients as one structured object and then regularize it jointly. As such, a covariate is selected consistently across the full day-ahead horizon, enabling the model to capture cross-hour feature group effects while maintaining model sparsity. 

Specifically, we collect the hourly coefficients into a single coefficient matrix,   $\mathbf{B}$, defined as:
\[
    \mathbf{B}
    =
    \bigl[\boldsymbol{\beta}_1,\dots,\boldsymbol{\beta}_{24}\bigr]
    \in \mathbb{R}^{M\times 24},
\]
where the $h$-th column $\boldsymbol{\beta}_h$ contains the regression coefficients for hour $h$. 
For each day $d$, the multivariate linear regression model adopted in CING-LEAR is expressed as in (\ref{eq:cing_lear_model}). 
\begin{equation}
    \mathbf{p}_d
    = \mathbf{B}^\top \mathbf{x}_d + \boldsymbol{\varepsilon}_d,
    \qquad d=1,\dots,N,
    \label{eq:cing_lear_model}
\end{equation}
where $\boldsymbol{\varepsilon}_d\in\mathbb{R}^{24}$ collects the hourly errors. Stacking all days, we collect the pricing vectors and error terms in the matrices $\mathbf{P}$ and $\mathbf{E}$, respectively, defined as: 
\[
    \mathbf{P}
    =
    \begin{bmatrix}
        \mathbf{p}_1^\top\\
        \vdots\\
        \mathbf{p}_N^\top
    \end{bmatrix}
    \in\mathbb{R}^{N\times 24},
\quad
    \mathbf{E}
    =
    \begin{bmatrix}
        \boldsymbol{\varepsilon}_1^\top\\
        \vdots\\
        \boldsymbol{\varepsilon}_N^\top
    \end{bmatrix}
    \in\mathbb{R}^{N\times 24}.
\]
Then, the formulation in \eqref{eq:cing_lear_model} can be written in matrix format as in (\ref{eq:cing_lear_matrix}). 
\begin{equation}
    \mathbf{P} = \mathbf{X}\mathbf{B} + \mathbf{E}.
    \label{eq:cing_lear_matrix}
\end{equation}

Let $\mathbf{b}_{j,\cdot}^\top$ denote the $j$-th row of $\mathbf{B}$, corresponding to the coefficients of feature $j$ across all 24 hours:
\[
    \mathbf{b}_{j,\cdot}
    =
    \bigl(\beta_{j,1},\dots,\beta_{j,24}\bigr)^\top \in\mathbb{R}^{24},
    \qquad j=1,\dots,M.
\]
To enforce joint feature selection across hours, we apply a group Lasso penalty on the rows of $\mathbf{B}$. Specifically, the penalty is defined as: 
\[
    \Omega(\mathbf{B})
    =
    \sum_{j=1}^{M}
    \Vert \mathbf{b}_{j,\cdot} \Vert_2,
\]
which encourages entire rows of $\mathbf{B}$ to shrink to zero. As a consequence, feature $j$ is either selected for all hours or removed simultaneously from the model.

Given $\mathbf{X}$ and $\mathbf{P}$, CING-LEAR solves the optimization problem expressed in (\ref{eq:cing_objective}). 
\begin{equation}
    \widehat{\mathbf{B}}
    =
    \arg\min_{\mathbf{B}\in\mathbb{R}^{M\times 24}}
    \Bigg\{
        \frac{1}{N}
        \Vert \mathbf{P}-\mathbf{X}\mathbf{B} \Vert_F^2
        + \lambda
\Omega(\mathbf{B})
    \Bigg\},
    \label{eq:cing_objective}
\end{equation}
where $\Vert\cdot\Vert_F$ denotes the Frobenius norm and $\lambda \ge 0$ is a regularization hyperparameter controlling the degree of sparsity.

Compared with LEAR, our formulation jointly estimates the full coefficient matrix $\mathbf{B}$, which results in a comparatively large set of parameters. In this case, the group Lasso enforces sparsity enabling consistent feature selection by exploiting the cross-hour group structure in electricity pricing signals. 

Given the fitted coefficient matrix $\hat{\mathbf{B}}$ obtained from (\ref{eq:cing_objective}), the day-ahead point forecast of price vector $\mathbf{p}_d \in \mathbb{R}^{24}$ can be defined as in (\ref{eq:cing_mean_forecast}). 
\begin{equation}
\widehat{\mathbf{p}}_d
=
\mathbb{E}(\mathbf{p}_d \mid \mathbf{x}_d)
=
\widehat{\mathbf{B}}^\top \mathbf{x}_d.
\label{eq:cing_mean_forecast}
\end{equation}
Uncertainty information can be derived by imposing a distributional assumption on the error vector. Here, we let $\boldsymbol{\epsilon}_d \sim \mathcal{N}(\mathbf{0}, \boldsymbol{\Sigma})$, such that $\boldsymbol{\Sigma} \in \mathbb{R}^{24 \times 24}$ is a symmetric positive
definite matrix that captures the dependence structure of forecast errors across different hours. The Gaussian assumption is non-problematic after an appropriate transformation described in Section \ref{sec:preprocess}. As such, the conditional distribution of the day-ahead price
vector given the feature vector $\mathbf{x}_d$ can be expressed as in (\ref{eq:distribution}).
\begin{equation}
\mathbf{p}_d \mid \mathbf{x}_d
\sim \mathcal{N}(\widehat{\mathbf{B}}^\top \mathbf{x}_d, \boldsymbol{\Sigma}).
\label{eq:distribution}
\end{equation}
The covariance matrix $\boldsymbol{\Sigma}$ can be estimated from in-sample
residuals using the empirical covariance estimator, as in (\ref{eq:cov}).  
\begin{equation} \label{eq:cov}
\widehat{\boldsymbol{\Sigma}}
=
\frac{1}{N}
\sum_{d=1}^N
\widehat{\boldsymbol{\epsilon}}_d
\widehat{\boldsymbol{\epsilon}}_d^\top,
\end{equation}
such that $\widehat{\boldsymbol{\epsilon}}_d
=
\mathbf{p}_d - \widehat{\mathbf{B}}^\top \mathbf{x}_d$. The resulting predictive distribution enables the construction of
useful uncertainty outputs, including prediction intervals and scenario
generation, while explicitly accounting for cross-hour error dependence.

For training and evaluation, we adopt a sliding window approach, with the evaluation period spanning two full years (2024 and 2025). When less than three years of historical observations are available, we use the longest training window supported by the available data.
We follow the training scheme proposed by \cite{Calibration_windows} where we combine forecasts generated by the same model trained over multiple ``calibration'' windows  
\citep{hubicka2018note}). Accordingly, a separate CING-LEAR model is trained using data from the most recent $N_c$ days preceding each forecast date, where $N_c$ denotes the calibration window length. The set of calibration windows considered includes four short-term windows: ${28,56,84,112}$ days (corresponding to $4$, $8$, $12$, and $16$ weeks, respectively), and three long-term windows of one, two, and three years, respectively. 
The resulting forecasts from all base models are combined adaptively through a 
weighted ensemble, where the ensemble weights are determined based on each model's recent predictive performance. 
Specifically, 
for a testing day $\ell$, the ensemble weight assigned to the $c$th base model is computed using its forecast error on the preceding day, as follows:
\begin{equation}
w_{\ell,c}
=
\frac{\overline{|\boldsymbol{\varepsilon}_{\ell-1,c}|}^{-1}}
{\sum_{j=1}^{C}\overline{|\boldsymbol{\varepsilon}_{\ell-1,j}|}^{-1}},
\label{eq:ensemble_weights}
\end{equation}
where $\overline{|\boldsymbol{\varepsilon}_{\ell-1,c}|}$ denotes the average absolute forecast error over the 24 hourly forecasts on the day preceding the testing day $\ell$, and $C$ is the number of calibration windows considered. 

Algorithm~\ref{alg:CING_LEAR} summarizes the implementation of CING-LEAR. For each base model $c$ and each testing day~$\ell$, the regularization parameter $\lambda^*_{\ell,c}$ is selected via five-fold cross-validation over $n_\lambda=100$ candidate values, with a maximum of $I_{\max}=5000$ iterations. The coefficient matrix $\widehat{\mathbf{B}}_{\ell,c}$ is then estimated by solving the optimization problem in (\ref{eq:cing_objective}) using coordinate descent algorithm implemented in \texttt{scikit-learn}. 
This approach is computationally efficient and particularly effective for sparse, multi-task regression problems \citep{Coordinate_descent}. The estimated coefficient matrix is then used to generate the day-ahead electricity forecasts, in hourly resolution. We then advance by one day, and the full procedure is repeated until the evaluation period is expended. 
\begin{algorithm}[]
\caption{Sliding Window Implementation of CING-LEAR}
\label{alg:CING_LEAR}
\begin{algorithmic}[1]

\State \textbf{Parameters:} test length $L$, CV folds $K$, grid size $n_{\lambda}$, max iterations $I_{\max}$,
\Statex \hspace{5.3em} calibration set, ensemble lookback $d$

\State Initialize the testing day index $\ell \leftarrow 1$

\While{$\ell \le L$}

    \For{$c = 1$ \textbf{to} $C$}

        \State Collect the previous $N_c$ days of data before day $\ell$
        \State Build training inputs and targets:
        $\mathbf{X}_{\ell,c} \leftarrow \{x_{d,h}\}_{d=1,\dots,N_c,\,h=1,\dots,24}$,
        \Statex \hspace{16.2em} $\mathbf{P}_{\ell,c} \leftarrow \{p_{d,h}\}_{d=1,\dots,N_c,\,h=1,\dots,24}$
        \State Build forecast inputs:
        \hspace{4.3em} $\mathbf{X}^{\mathrm{fcst}}_{\ell,c} \leftarrow \{x_{d,h}\}_{d=1,\dots,N_c+1,\,h=1,\dots,24}$

        \State Select $\lambda^*_{\ell,c} \leftarrow CV(K,n_{\lambda},I_{\max})$

        \State $\widehat{\mathbf{B}}_{\ell,c} \leftarrow \text{CING-LEAR}(\mathbf{X}_{\ell,c},\mathbf{P}_{\ell,c};\lambda^*_{\ell,c},I_{\max})$, $\widehat{\mathbf{P}}_{\ell,c} \leftarrow \mathbf{X}^{\mathrm{fcst}}_{\ell,c}\,\widehat{\mathbf{B}}_{\ell,c}$

    \EndFor

    \If{$\ell \le d$} $\widehat{\mathbf{P}}^{\mathrm{ens}}_{\ell} \leftarrow \frac{1}{C}\sum_{c=1}^{C}\widehat{\mathbf{P}}_{\ell,c}$
    \Else
        \For{$c = 1$ \textbf{to} $C$}
            \State Compute      $\overline{|\boldsymbol{\varepsilon}_{\ell-1,c}|}$.
        \EndFor
        \State Compute $\{w_{\ell,c}\}_{c=1}^{C}$ using~(\ref{eq:ensemble_weights}), and construct the ensemble $\widehat{\mathbf{P}}^{\mathrm{ens}}_{\ell} \leftarrow \sum_{c=1}^{C} w_{\ell,c}\,\widehat{\mathbf{P}}_{\ell,c}$
    \EndIf

    \State $\ell \leftarrow \ell + 1$

\EndWhile

\State \Return $\{\widehat{\mathbf{B}}_{\ell,c}\}$, $\{\widehat{\mathbf{P}}_{\ell,c}\}$, $\{\widehat{\mathbf{P}}^{\mathrm{ens}}_{\ell}\}$

\end{algorithmic}
\end{algorithm}

\section{Experimental Setup}
\label{sec:experiment}
In this section, we describe data preprocessing steps, benchmarks, and metrics used to evaluate CING-LEAR. 

\subsection{Data preprocessing}
\label{sec:preprocess}
To handle heavy tails and price spikes, we pre-process electricity prices in two steps: a normalization and a transformation, as recommended in \cite{Bartosz} and \cite{ lago2021forecasting}. 
Let \(p_{d,h}\) denote the price at day \(d\) and hour \(h\). The normalized price is defined as:
\begin{equation}
\tilde{p}_{d,h} = \frac{p_{d,h} - a}{b},
\end{equation}
where \(a\) is the median of the in-sample prices, and \(b\) is the sample median absolute deviation (MAD) adjusted by a factor \(1.4826 \approx 1/z_{0.75}\) where \(z_{0.75}\) is the 75th percentile of the standard normal distribution. 
After normalization, we apply the inverse hyperbolic sine transformation. The transformed price is given by:
\begin{equation}
\operatorname{asinh}(\tilde{p}_{d,h})
= \log \left( \tilde{p}_{d,h} + \sqrt{\tilde{p}_{d,h}^{2} + 1} \right).
\end{equation}

\subsection{Benchmarks}
\label{sec:benchmarks}
Several representative benchmarks in EPF studies are considered. 
All benchmark models are implemented 
under the same sliding window setup as in CING-LEAR. 
Probabilistic forecasts are constructed using the same approach described in Section \ref{sec:CING}. The benchmarks, denoted as $\mathcal{B}1$-$\mathcal{B}7$, are described as follows: 
\begin{enumerate}[label=({$\mathcal{B}$}{{\arabic*}})]
\item \textbf{Na\"ive:} The Na\"ive approach is a standard benchmark in the forecasting literature and practice and simply assumes persistence of current LMPs to the following day, as in (\ref{eq:naive}). 
\begin{equation}
\label{eq:naive}
\hat{p}_{d,h}^{\text{Na\"ive}} = p_{d-1,24}, 
\quad \forall h \in \{1,\dots,24\}.
\end{equation}

\item \textbf{Seasonal Na\"ive:}
The Seasonal Na\"ive extends the na\"ive model by accounting for daily seasonality, 
as in (\ref{eq:seasonal_naive}). 
\begin{equation}
\label{eq:seasonal_naive}
\hat{p}_{d,h}^{\text{S-Na\"ive}} = p_{d-1,h},
\quad \forall h \in \{1,\dots,24\}.
\end{equation}

\item \textbf{ARIMA:}
The autoregressive integrated moving average (ARIMA) model is a univariate time series approach 
which linearly extrapolates historical signals, as in (\ref{eq:arima}). 
\begin{equation}
\label{eq:arima}
\begin{aligned}
\hat{p}_{d,h}^{\text{ARIMA}}
= \phi_0
+ \sum_{i=1}^{p} \phi_i p_{d,h-i}
+ \sum_{j=1}^{q} \theta_j \varepsilon_{d,h-j} + \varepsilon_{d,h},
\qquad \forall h \in \{1,\dots,24\}.
\end{aligned}
\end{equation}

\item \textbf{ARIMA-X:}
The ARIMA-X includes exogenous variables in the ARIMA structure, as in (\ref{eq:arimax}). 
\begin{equation}
\label{eq:arimax}
\begin{aligned}
\hat{p}_{d,h}^{\text{ARIMA-X}}
={}& \phi_0
+ \sum_{i=1}^{p}\phi_i p_{d,h-i}
+ \sum_{j=1}^{q}\theta_j\varepsilon_{d,h-j}
+ \sum_{k=1}^{K}\beta_k x_{d,h}^{(k)}
+ \varepsilon_{d,h},
\qquad \forall h\in\{1,\dots,24\}.
\end{aligned}
\end{equation}
where \( x_{d,h}^{(k)} \) represents the day-ahead forecast of the \(k\)-th exogenous variable. For a fair comparison, ARIMA-X uses the same exogenous information as CING-LEAR, including forecasts of regional load, solar generation, natural gas generation, and fuel prices.



\item \textbf{LEAR:}
The Lasso-Estimated Autoregressive (LEAR) model serves as the univariate counterpart of CING-LEAR, providing a natural baseline to assess the value of multivariate cross-output learning and group regularization. It is presented in Section~\ref{sec:lear}. More details can be found in~\cite{uniejewski2016automated}. For a fair comparison, LEAR is likewise trained using multiple calibration windows, and its forecasts are combined using the same adaptive ensemble procedure described in Section ~\ref{sec:CING}. Due to its larger feature set, only longer calibration windows are considered, namely three two-year windows (714, 721, and 728 days) and three three-year windows (1078, 1085, and 1092 days). The implementation details closely follow the recommendations in \cite{lago2021forecasting}.

\item \textbf{DNN:}
The Deep Neural Network (DNN) benchmark, as proposed in \cite{lago2018forecasting}, is a fully connected multi-layer perceptron (MLP) with a multi-output structure (i.e., 24 output nodes) for jointly predicting the vector of hourly day-ahead LMPs. The architecture consists of four layers: an input layer with the same inputs as in CING-LEAR, two hidden layers, and a linear output layer with 24 outputs representing the hourly LMP of the day. Each hidden layer is fully connected and employs the ReLU activation function. The output layer uses a linear activation to generate point forecasts. 
The implementation details and hyperparameter settings closely follow the recommendations in \cite{lago2021forecasting}. 


\item \textbf{Chronos-2:} Chronos-2 is a pre-trained time series foundation model (TSFM) \citep{Chronos-2}. 
It leverages in-context learning and produces multi-step quantile 
forecasts. 
We include two variants of Chronos-2: a \textit{zero-shot} variant in which the model is applied without task-specific parameter fine-tuning; and a \textit{fine-tuned} variant wherein 
we set the context length to 2048 (maximum setting) with a 24-hour 
prediction length. Our choice of Chronos-2 as a representative of the TSFM paradigm is motivated by its strong performance in recent EPF benchmarking studies  \citep{pan2026benchmarking, pan2026review}. 

\end{enumerate}

\subsection{Evaluation metrics}
\label{sec:evaluation}
Three metrics are used to evaluate the predictive performance of CING-LEAR. For point forecast evaluation,     
we use the Mean Absolute Error (MAE) and the Root Mean Squared Error (RMSE). MAE is defined as in \eqref{eq:mae}, where \(p_{d,h}\) denotes the observed electricity price at day \(d\) and hour \(h\), \(\hat{p}_{d,h}\) is the corresponding point forecast, and \(N\) is the number of days in the test set.
\begin{equation} \label{eq:mae}
\mathrm{MAE} = \frac{1}{24\cdot N}\sum_{d=1}^{N}\sum_{h=1}^{24}\left|p_{d,h} - \hat{p}_{d,h}\right|,
\end{equation}
RMSE is defined as in (\ref{eq:rmse}), penalizing large forecast errors more heavily by means of its quadratic loss, thereby providing complementary information to MAE regarding the dispersion of point forecast errors.
\begin{equation} \label{eq:rmse}
\mathrm{RMSE} = \sqrt{\frac{1}{24\cdot N}\sum_{d=1}^{N}\sum_{h=1}^{24}\left(p_{d,h} - \hat{p}_{d,h}\right)^2},
\end{equation}

For probabilistic forecast evaluation, we use the Continuous ranked probability score (CRPS). For a predictive distribution \(F_{d,h}(\cdot)\) and an observation \(p_{d,h}\), the CRPS is defined as
\begin{equation}
\begin{aligned}
\operatorname{CRPS}(F_{d,h},p_{d,h})
 =
\int_{-\infty}^{+\infty}
\left(
F_{d,h}(z)-\mathbb{I}\{z\ge p_{d,h}\}
\right)^2\,\mathrm{d}z .
\end{aligned}
\end{equation}
where \(\mathbb{I}(\cdot)\) denotes the indicator function. 
We compute CRPS using a sample-based approximation. Given \(M\) Monte Carlo samples \(\{x_{d,h}^{(m)}\}_{m=1}^{M}\) drawn from the predictive distribution \(F_{d,h}\), where \(M\) is chosen to be sufficiently large ($M=20{,}000$ in this study), CRPS can be expressed 
as:
\begin{equation}
\begin{aligned}
\operatorname{CRPS}(F_{d,h},p_{d,h})
\approx
\widehat{\operatorname{CRPS}}(F_{d,h},p_{d,h})
=
\frac{1}{M}\sum_{m=1}^{M}
\left|x_{d,h}^{(m)}-p_{d,h}\right|
-\frac{1}{2M^2}
\sum_{m=1}^{M}\sum_{m'=1}^{M}
\left|x_{d,h}^{(m)}-x_{d,h}^{(m')}\right|.
\end{aligned}
\end{equation}
The average CRPS over the testing period is then computed as:
\begin{equation}
\mathrm{CRPS} \approx \frac{1}{24\cdot N}\sum_{d=1}^{N}\sum_{h=1}^{24}
\widehat{\mathrm{CRPS}}(F_{d,h}, p_{d,h}).
\end{equation}

\section{Results and Analysis}
\label{sec:results}
We first present the evaluation results on two years of CAISO data in Section \ref{sec:fores}. This is followed by an examination of feature recovery in Section \ref{sec:feature selection}. Sections \ref{sec:pge} and \ref{sec:industry} evaluate CING-LEAR on a recently concluded forecasting competition, and independently against two operational industry benchmarks in CAISO, respectively. 

\subsection{Forecasting results on two years of CAISO data} \label{sec:fores}
\begin{figure*}
    \centering
    \includegraphics[trim = 0 0.75cm 0 0, width=1\linewidth]{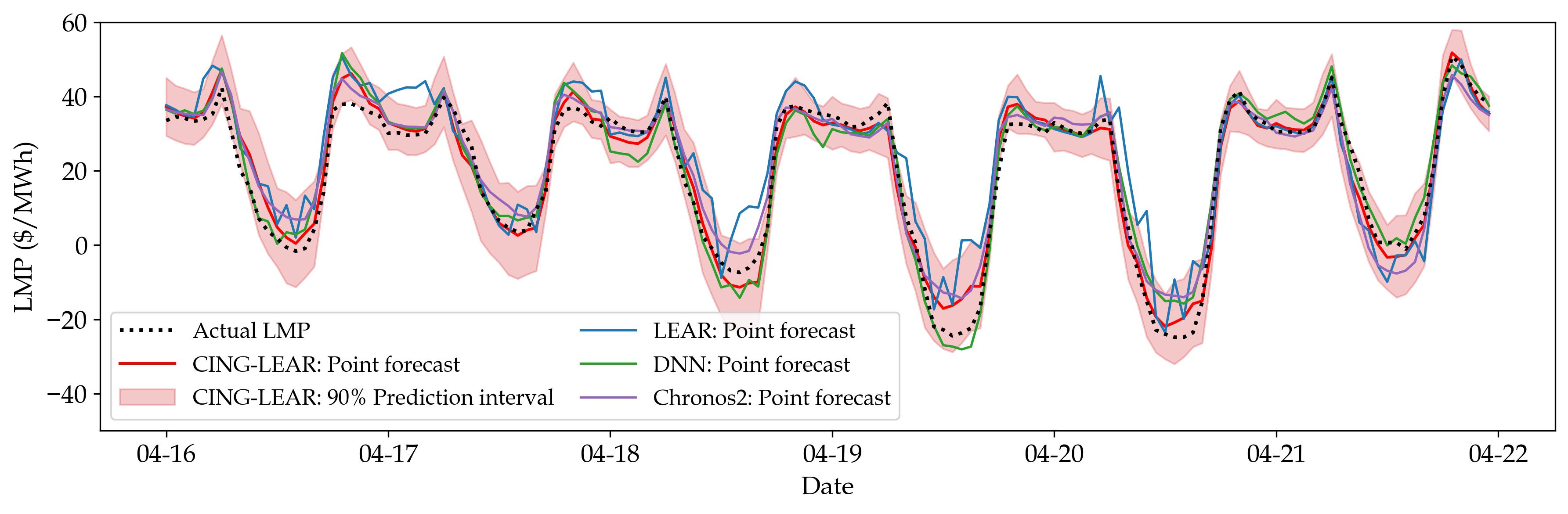}
    \caption{Comparison of actual LMPs and the corresponding day-ahead forecasts over a representative period of one week in April 2025. The dashed line denotes the actual LMPs, while solid lines represent the forecasts produced by the different models (red = CING-LEAR, blue = LEAR, green = DNN, purple = Chronos-2). The red shaded region indicates the 90\% prediction interval produced by CING-LEAR, showing faithful probabilistic coverage.}
    \label{fig:Comparison}
\end{figure*}

Figure~\ref{fig:Comparison} presents a representative period in April $2025$, comparing the actual LMPs with the corresponding forecasts generated by LEAR, DNN, Chronos-2, and CING-LEAR. All four models appear to successfully capture the daily seasonality of LMPs, with lower prices during daytime hours and higher prices overnight, as well as two peaks around sunrise and sunset. 
For this representative period, all four models 
achieve relatively higher accuracy during nighttime. 
This may be caused by higher load and solar penetration during the day, which collectively induce greater variability into LMPs. A noticeable visual distinction is how CING-LEAR produces smoother forecasts than LEAR, which is due to the grouped feature selection. 
The temporal smoothness of CING-LEAR's forecasts 
appear to align with actual LMP patterns, leading to improved accuracy and more realistic behavior (see, for example, 
the forecasts on April $20$). 

\begin{table*}[htbp]
\centering
\caption{Performance comparison among different forecasting models in terms of MAE, RMSE, and CRPS. Bold-faced 
values indicate the top 
performance.} 
\label{tab:model_mae_rmse_crps}
\renewcommand{\arraystretch}{1}
\begin{adjustbox}{max width=\textwidth}
\begin{tabular}{|c|l|ccc|ccc|ccc|}
\hline
& \multirow{2}{*}{Method}
& \multicolumn{3}{c|}{MAE}
& \multicolumn{3}{c|}{RMSE}
& \multicolumn{3}{c|}{CRPS} \\
\cline{3-11}
&   
& 2024 & 2025 & Overall
& 2024 & 2025 & Overall
& 2024 & 2025 & Overall
\\
\hline
\multirow{6}{*}{
\makecell{Time Series \\ Benchmarks}
}
& Na\"ive      
& $10.670$& $9.734$& $10.203$
& $20.593$& $14.433$& $17.786$& $8.026$& $6.634$& $7.330$\\

& Seasonal Na\"ive    
& $7.477$& $5.791$& $6.635$
& $15.892$& $8.491$& $12.745$& $6.010$& $4.257$& $5.133$\\

& ARIMA      
& $10.595$& $9.592$& $10.094$& $20.687$& $14.176$& $17.737$& $8.006$& $6.579$& $7.292$\\

& ARIMA (w/ transform)
& $10.546$& $9.607$& $10.077$& $20.856$& $14.211$& $17.850$& $7.986$& $6.584$& $7.285$\\

& ARIMA-X    
& $8.057$& $5.274$& $6.667$
& $16.903$& $7.128$& $12.978$& $6.504$& $3.822$& $5.163$\\

& ARIMA-X (w/ transform)
& $7.001$& $5.264$& $6.134$
& $14.960$& $7.361$& $11.794$& $5.641$& $3.845$& $4.743$\\

\hline
\multirow{4}{*}{\makecell{Advanced \\ Methods}}
& LEAR \citep{uniejewski2016automated})
& $6.309$& $4.490$& $5.399$& $13.968$& $6.313$& $10.844$& $5.025$& $3.251$&$4.138$\\

& DNN  \citep{lago2018forecasting})     
& $6.245$& $4.673$& $5.460$& $\bf{13.002}$& $6.502$& ${10.284}$& ${4.787}$ & $3.388$ & $4.088$ \\

& Chronos-2 (zero-shot) \citep{Chronos-2})
& $6.058$& $4.096$& $5.078$& $14.104$& $5.730$& $10.770$& $5.009$& $2.996$& $4.002$\\

& Chronos-2 (fine-tuned) \citep{Chronos-2})
& ${5.784}$& ${3.946}$& ${4.866}$& $13.618$& ${5.583}$& $10.412$& ${4.787}$& ${2.900}$& ${3.844}$\\

\hline
\multirow{2}{*}{Proposed}
& CING-LEAR
& $5.875$& $4.184$& $5.031$& $13.805$& $6.013$& $10.653$& $4.797$& $3.064$& $3.930$\\


& CING-LEAR + Chronos-2 (fine-tuned)
& $\bf{5.709}$& $\bf{3.875}$& $\bf{4.793}$& ${13.259}$& $\bf{5.435}$& $\bf{10.138}$& $\bf{4.671}$& $\bf{2.829}$&$\bf{3.750}$\\

\hline
\end{tabular}
\end{adjustbox}
\end{table*}

Table~\ref{tab:model_mae_rmse_crps} reports the average performance for $2024$ and $2025$, as well as the two-year average, across all methods considered. \textit{Transform} refers to whether the data preprocessing steps described in Section~\ref{sec:preprocess} were applied. Please note that both LEAR and CING-LEAR use this transformation as well. 
Several observations can be drawn from Table \ref{tab:model_mae_rmse_crps}. \textit{First,} both LEAR and DNN clearly outperform the standard time series benchmarks\textemdash a conclusion that supports the potency of these methods 
for short-term EPF, as similarly demonstrated in prior studies \citep{lago2021forecasting}. 
Second, CING-LEAR consistently outperforms LEAR and DNN (with the exception of RMSE and CRPS in 2024, where DNN is leading). This is re-assuring since both LEAR and DNN are well-proven methods that have been tailored for short-term EPF applications. 

Compared with the zero-shot Chronos-2, CING-LEAR achieves a modest but meaningful two-year average improvement of approximately $1\%-2\%$, although the relative performance varies by year. Specifically, CING-LEAR outperforms Chronos-2 in 2024 by $\sim$3\%, 2\%, and 4\% in terms of MAE, RMSE, and CRPS, respectively, whereas Chronos-2 improves in 2025 by $\sim$2\%, 5\%, and 2\%, respectively. Fine-tuned Chronos-2 improves upon its zero-shot counterpart and subsequently maintains a small but consistent advantage over CING-LEAR, with average gains of $\sim2\%-3\%$. 

The comparison between CING-LEAR and Chronos-2 should be interpreted in light of their fundamentally different learning paradigms. Chronos-2 benefits from large-scale pretraining on diverse time series followed by task-specific fine-tuning, whereas CING-LEAR is exclusively locally trained. Despite this distinction, CING-LEAR achieves better performance than the zero-shot TSFM and remains within a narrow margin of the fine-tuned version. These findings suggest that carefully designed domain-specific forecasting models remain highly competitive, even when compared with state-of-the-art pretrained TSFMs. Beyond predictive accuracy, practical deployment involves additional considerations. In our experiments, fine-tuning Chronos-2 requires $\sim30$ minutes per model update, whereas training CING-LEAR requires only $\sim3-4$ minutes. Although this difference may be modest for a single forecasting task, it becomes increasingly important when maintaining forecasting models across large electricity markets containing hundreds or thousands of pricing nodes. 

Another equally important aspect is explainability. CING-LEAR is intrinsically interpretable, allowing forecast users to directly examine the contributions of 
exogenous variables, diagnose abnormal predictions, and probe the market and system conditions driving forecast behavior (we dedicate Section \ref{sec:feature selection} for this aspect). 
In contrast, TSFMs are generally perceived as black-boxes, often requiring post-hoc analysis for explainability. 

Importantly, the comparison between CING-LEAR and 
TSFMs reveals that 
these two fundamentally different model architectures can exhibit complementary strengths.  
As such, we consider a simple ensemble constructed as the mean of the predictions from CING-LEAR and Chronos-2 (fine-tuned). As shown in Table~\ref{tab:model_mae_rmse_crps}, the ensemble achieves a significantly better performance over its member constituents, 
suggesting that ``the whole may be greater than the sum of its parts.''
This result indicates that CING-LEAR captures predictive information that complements, rather than overlaps with, the representations learned by TSFMs. More broadly, it suggests that the future of electricity price forecasting may lie not in replacing domain-tailored forecasting methods with TSFMs, but potentially in combining their complementary strengths for improved forecast accuracy. This conclusion aligns with recent findings in the EPF literature \citep{pan2026benchmarking, pan2026review}.  

\subsection{Feature selection}
\label{sec:feature selection}
\begin{figure*}
    \centering
    \includegraphics[width=1\linewidth]{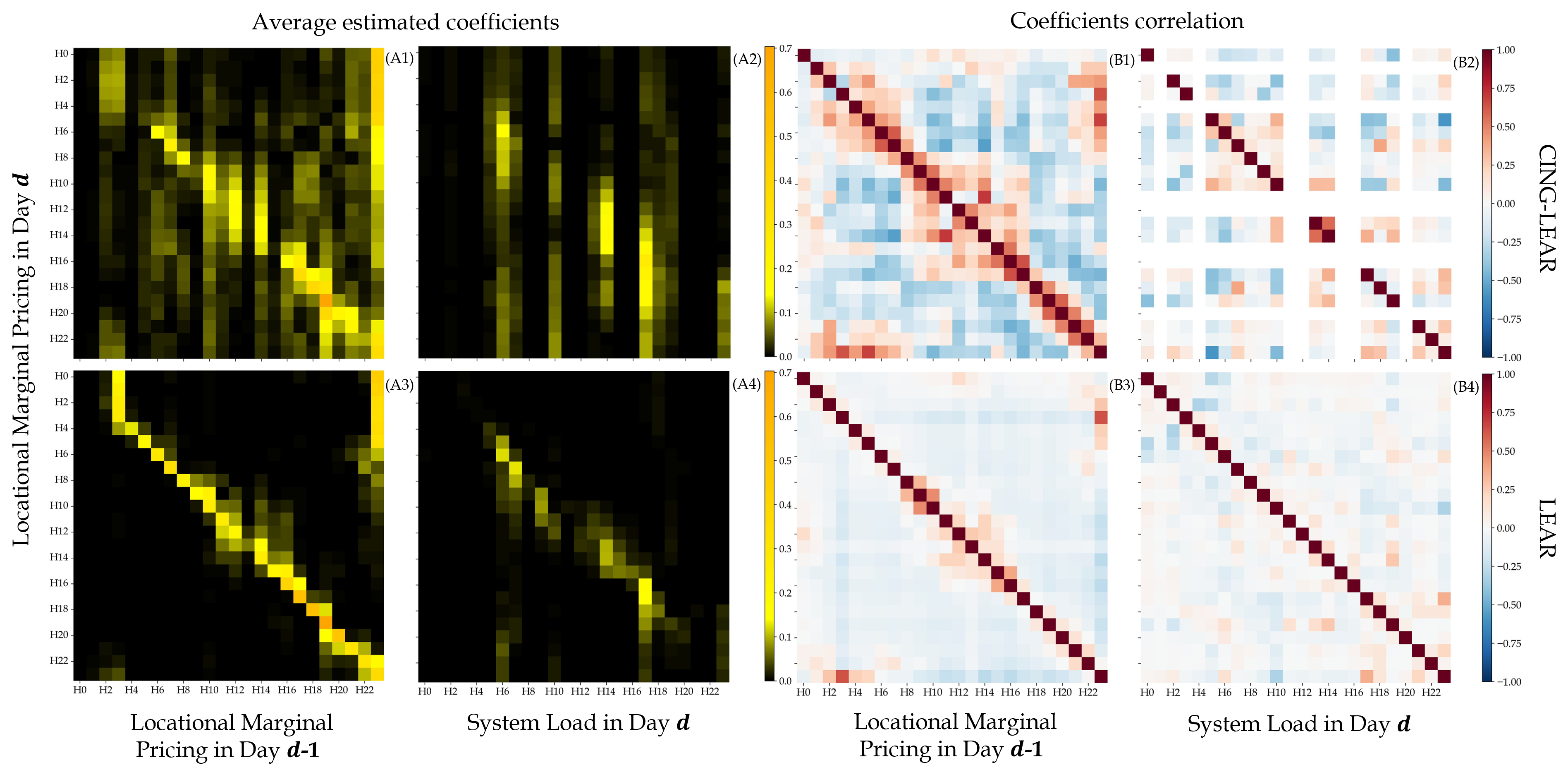}
    \caption{The left set of panels (A1-A4) displays the estimated coefficients of each feature across the 24 hours, averaged over the entire test period, where black denotes zero coefficients and shades of yellow represent absolute coefficient magnitudes. The right set of panel (B1-B4) shows the correlation matrices of the coefficients across the 24 hours, where red and blue denote positive and negative correlations, respectively. The first and second rows corresponds to CING-LEAR and LEAR, respectively, both calibrated with a window of 1092 days. Due to page limitations, only two representative features are shown: the historical price from the previous day $d-1$ and the load forecast for the current day $d$. }
    \label{fig:combined_figure}
\end{figure*}

Next, we probe CING-LEAR's ability to capture realistic group effects, by examining the set of its estimated coefficients. Figure~\ref{fig:combined_figure}(A) compares the estimated coefficients for CING-LEAR (top) and LEAR (bottom), both calibrated with a window of $1092$ days. The LEAR coefficients are largely concentrated around the diagonal, indicating that for each hourly LMP, only features associated with the same or nearby hours are retained. In contrast, CING-LEAR selects a broader set of features by capturing temporal group effects, aligning with the empirical analysis of pricing signals in Figure \ref{fig:correlation}. 
Interestingly, CING-LEAR also reinforces certain informative features already identified by LEAR. For instance, the LMP at the last hour in Day $d-1$ and the load at hours $14$ and $17$ show larger coefficient magnitudes under CING-LEAR than in LEAR. 

Figure \ref{fig:combined_figure}(B) shows the correlations among the estimated coefficients from CING-LEAR (top) and LEAR (bottom), 
wherein CING-LEAR exhibits more pronounced correlations among its coefficients across the day, indicating a stronger cross-hour structure in feature effects. Visually, 
the correlation matrix from CING-LEAR can be partitioned into three segments: hours $0$- $7$, $8$-$17$, and $18$-$23$. These segments appear to maintain highly positive in-segment correlations, and negative between-segment correlations. 
The relative smoothness of CING-LEAR's forecasts (shown in Figure \ref{fig:Comparison}) can be explained in light of these pronounced correlations.

\subsection{Benchmarking against the 2024 PG\&E Energy Analytics competition} \label{sec:pge}
We further evaluate the performance of CING-LEAR by comparing it with 
the top performing teams in the 2024 PG\&E Energy Analytics Challenge 2024 that focused on short-term EPF 
\citep{IISE_PGE}. The training dataset consists of three years of hourly LMPs for the NP15 region, from 2020 to 2022. A limited set of exogenous variables is provided, including regional load and natural gas prices.  
The competition allowed participants to incorporate supplementary exogenous information into their forecasting models. The evaluation was carried out on a reserved test set that included LMPs from the first two weeks of 2023. During the forecasting stage, future day-ahead LMPs are excluded from the model inputs, and only historical exogenous information available up to the forecast time is permitted. 

To ensure compliance with the competition rules, 
the day-ahead LMP forecasts for each day are used as inputs for subsequent daily forecasts throughout the two-week evaluation period. 
In Table \ref{tab:IISE_PGE}, we compare the performance of CING-LEAR calibrated with a window of $1092$ days, relative to the three top-ranked teams from the challenge, along with the performance of statistical benchmarks as reported by the challenge organizers \citep{IISE_PGE}. 
The results show that Team 1 (an LSTM model with exogenous features \citep{salihoglu2026adaptive}) consistently outperforms all competing approaches. The approach uses a custom loss function and benefits from the use of a large set of exogenous inputs (46 features). Without relying on additional features, CING-LEAR achieves second place among all participants. LEAR ranks a close third, further confirming its strength as a competitive model for short-term EPF.


\begin{table*}[htbp]
\centering
\caption{Performance comparison of CING-LEAR calibrated with a window of 1092 days, versus top-performing teams in the 2024 PG\&E Energy Analytics Competition, as reported by \cite{IISE_PGE}. 
\# of added features represent the additional covariates used, beyond those provided by the competition organizers. Bold-faced and underlined values indicate the top and second-best performance, respectively.}
\label{tab:IISE_PGE}
\renewcommand{\arraystretch}{1}
\begin{tabular}{|l|c|cc|}
\hline
Method & \makecell[l]{\# of added features} & MAE & RMSE \\
\hline

\makecell[l]{Team 1 (LSTM with exogenous features)} & 46 & $\bf{18.92}$ & $\bf{24.39} $\\

\makecell[l]{Team 2 (Tree-based ensemble)}          & $4$  & $29.08$ & $36.03$ \\

\makecell[l]{Team 3 (VAR + XGBoost)}                & $1$  & $26.82$ & $35.17$ \\

Na\"ive                               & $0$ & $41.01$ & $50.36$ \\
\makecell[l]{Seasonal Na\"ive (period = 24 hours)}  & $0$ & $38.00$ & $46.30$ \\
\makecell[l]{ARIMA (no exogenous features)}         & $0$ & $38.40$ & $48.10$ \\
\makecell[l]{ARIMA-X (w/ exogenous features)}       & $0$ & $32.30$ & $44.00$ \\
\hline

LEAR (\citep{uniejewski2016automated}) & $0$ & $27.78$ & $36.49$ \\

CING-LEAR (Proposed)                             & 0 & \underline{$23.61$} & \underline{$32.38$} \\
\hline

\end{tabular}
\end{table*}

\subsection{Benchmarking against industry operational models} \label{sec:industry}
CING-LEAR is independently validated against two 
models that are currently deployed by a major U.S. utility. Those are denoted as Forecast\_1 and Forecast\_2, respectively, and their technical details and identities are withheld. The evaluation was conducted independently: we first generated and submitted CING-LEAR forecasts without access to the forecasts produced by Forecast\_1 and Forecast\_2, and only subsequently obtained their forecasts to conduct the comparison. 

Table~\ref{tab:Industry} summarizes the forecasting performance of these two models and that of CING-LEAR 
over both 7-month (May to December 2024) and 12-month (January to December 2024) test periods. 
Over the 7-month period, where both industrial benchmarks are available, CING-LEAR achieves consistently competitive performance, ranking first in MAE and second in RMSE. Over the full 12-month period, which provides a more comprehensive evaluation, CING-LEAR outperforms the available benchmark, achieving improvements of approximately 17\% in MAE and 5\% in RMSE.
While the comparison set is more limited in this case, the results indicate strong and stable performance over an extended evaluation period. Overall, these findings demonstrate that CING-LEAR provides robust performance under strong competition and maintains high accuracy over both time spans, providing a strong indicator of industrial relevance, particularly given the maturity of such systems in practice.

\begin{table}[htbp]
\centering
\caption{Forecasting performance comparison with industrial benchmarks. Bold-faced and underlined values indicate the top and second-best performance, respectively.}
\label{tab:Industry}
\renewcommand{\arraystretch}{1.1}
\begin{tabular}{|l|cc|cc|}
\hline
&  \multicolumn{2}{c|}{7-month}    &   \multicolumn{2}{c|}{12-month}    \\
\hline
Method & MAE & RMSE  &  MAE &   RMSE\\
\hline
Forecast\_1 & $\underline{4.95}$& $12.82$  &   -   &   -  \\
Forecast\_2 & $5.84$ & {$\bf{11.93}$} &   \underline{$7.10$} &   \underline{$14.49$}   \\
\hline
CING-LEAR   & $\bf{4.85}$& \underline{$12.26$}&   $\bf{5.88}$&   $\bf{13.81}$\\

\hline
\end{tabular}
\end{table}

\section{Conclusion}
\label{sec:conclusion}
This paper proposes CING-LEAR, a multivariate statistical method for short-term electricity price forecasting. 
Extensive, multi-year evaluations 
demonstrate that CING-LEAR delivers consistently strong forecasting performance relative to established benchmarks. 
In addition to improvements in point and probabilistic forecast quality, the resulting forecasts exhibit smooth temporal profiles and interpretable signatures that reflect the model’s ability to capture realistic cross-hour dependencies in electricity pricing signals.

This study also highlights several promising directions for future research. First, our study shows that combinations of diverse forecasters (e.g., CING-LEAR with a TSFM) can indeed be promising. 
More advanced ensemble strategies could account for when and under which conditions these models perform better, for example based on performance, volatility, or regime indicators. Second, the proposed multivariate framework can be extended to jointly model electricity prices across multiple locations, enabling the explicit incorporation of spatial dependencies inherent in electricity pricing signals. 

\section*{Acknowledgment}
This research is supported in part by the U.S. National Science Foundation, ECCS \# 2114422. 

\bibliographystyle{unsrtnat}

\bibliography{cas-refs}


\end{document}